# The breadth premium: measuring the firm-level impact of CEO career breadth

PREPRINT VERSION 0.4
EARLIER RESULTS WITHDRAWN, DATABASE REBUILT, NEW RESULTS PUBLISHED IN VERSION 1.0

T. Alexander Puutio*
Faculty of Arts and Sciences
University Hall, Harvard Yard
Cambridge, MA, 02138
teemu_puutio@fas.harvard.edu

July 31, 2026

**ABSTRACT**

Prevailing career and education systems continue to reward early specialization and deep expertise within narrow domains. While such depth promotes efficiency, it may also limit adaptability in complex and rapidly changing environments. Building on research showing that variability in training inputs enhances learning outcomes across cognitive and behavioral domains, this study explores whether the same principle applies to executive performance.

These early findings suggest that leadership breadth, defined as experience spanning multiple functions, disciplines, and sectors, is positively associated with firm-level performance. While the dataset remains under validation, the pattern observed supports the emerging view that as specialization deepens, the marginal value of lateral insight rises. Breadth, in this light, functions as a form of adaptive capital: it enhances leaders' capacity for integrative reasoning, organizational translation, and strategic flexibility in uncertain environments.

## 1 Introduction

Careers are often described as paths, evoking images of winding journeys or singular pursuits. But beneath that metaphor lies a more structured reality where careers unfold as trajectories and predictable arcs that progress through a sequence of stages, each typically reinforcing deeper specialization within a given domain instead of inviting cross-discipline exploration. Classic career development theory has long emphasized this progression, most notably in Super's (1980) career lifespan framework, which outlines four key stages: exploration, establishment, maintenance, and disengagement. As individuals move through these stages, their work typically becomes more specialized, reinforcing a narrow but deep expertise within a particular domain.

These stages have been widely adapted in studies of professional and academic careers alike, and they imply a natural arc of increasing specialization where early career is a period of search and fit; mid-career is marked by consolidation and performance within a niche; and late career often brings a plateau or narrowing of scope as roles stabilize and productivity tapers (Savickas, 2002, Arthur, Hall, & Lawrence, 1989; Lent & Brown, 2013; Zacher et al., 2019). The structure of many professions reinforces this deepening arc. For example, in academia, career progression is codified into tiered systems from doctoral training and postdoctoral fellowships to assistant, associate, and full professorships within a single domain (Mantai & Marrone, 2023). Each tier increasingly rewards deeper expertise within a subfield rather than wider exploration across domains (Frølich et al., 2018; Fumasoli, Goastellec, & Kehm, 2015). Though country-specific differences exist in how career tiers are structured, the underlying logic of career progression remains largely consistent. As individuals advance, their specialization deepens, and their ability to transition between domains narrows (Coppola & Young, 2022; Sullivan & Crocitto, 2007). The result is a career funnel which begins with broad exposure to varied knowledge domains that progressively narrows into highly specific expertise (Figure 1). This progression can be conceptualized as a dynamic of "out-category pruning" i.e. the shedding of entire domains, and "in-category refining" where the focus is tightened within the new firmly established single discipline.

**Figure 1: The career funnel**

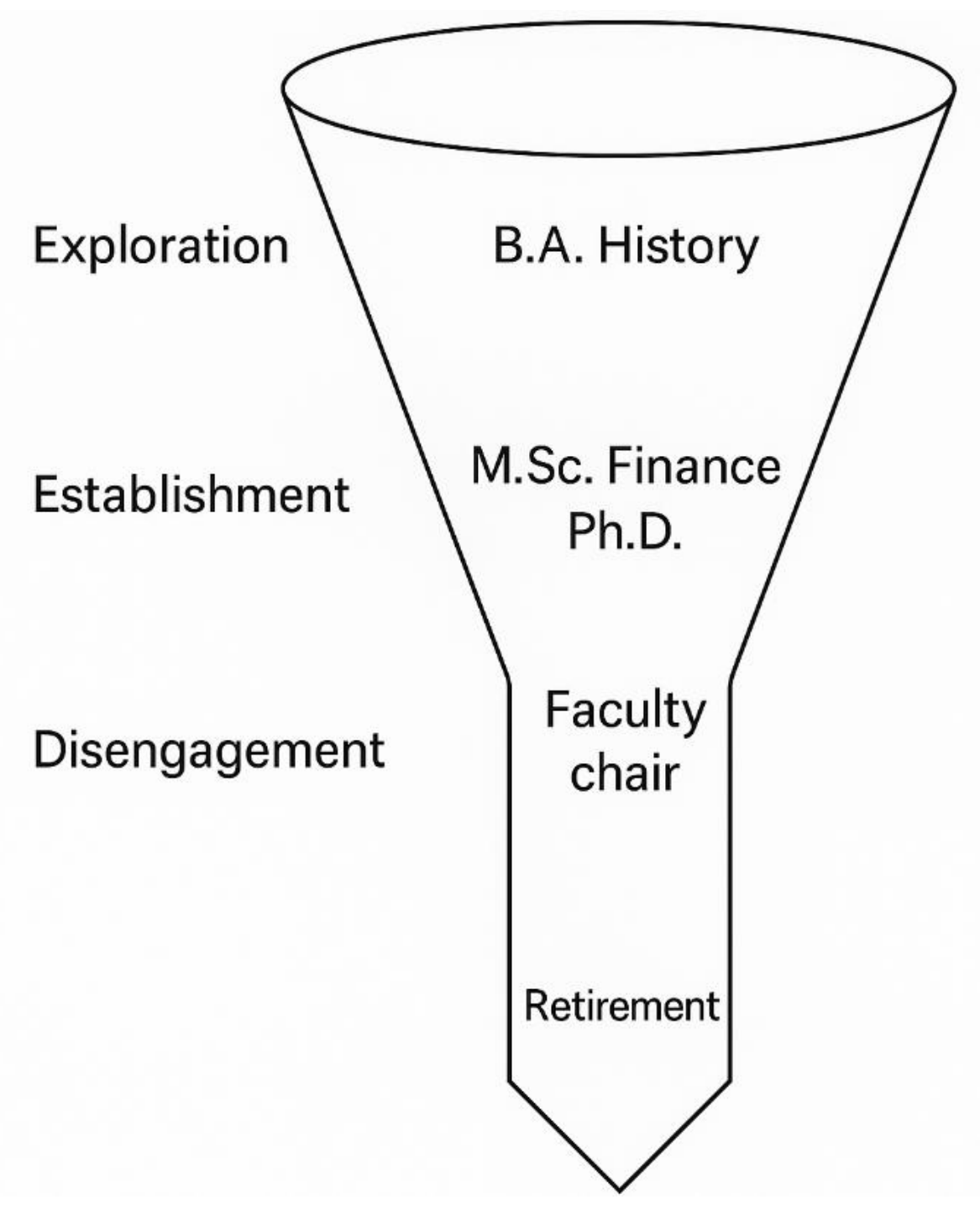

For example, an academic might begin with a B.A. in History, shift to an M.Sc. in Finance, and ultimately pursue a Ph.D. in Quantitative Economics. Over time, their life's work may focus on a tightly bounded subfield, with reduced engagement in adjacent disciplines.

This deepening specialization is neither accidental nor arbitrary. Instead, it is an adaptive and often necessary response to the increasing complexity of knowledge systems. As fields expand and professional standards rise, the entry threshold to participate meaningfully grows steeper. Abbott (1988) describes how professions construct and defend jurisdictional boundaries by creating specialized knowledge silos, which in turn require aspirants to invest more time in credentialing. Consequently, early specialization becomes a rational response to these structural incentives. Bidwell and Mollick (2015) further note that job descriptions and hiring practices have evolved to favor narrowly defined skill sets that reward internal mobility, especially in knowledge-intensive industries, reinforcing a preference for depth over breadth. Cappelli (2015) similarly observes that rising credential requirements and tighter skill matching have contributed to an increasingly overeducated labor force, where more individuals acquire specialized qualifications than many roles demand, generating inefficiencies for the system as a whole. Taken together, these patterns help explain how the career funnel emerges. As individuals respond to rising specialization demands, and firms create increasingly complex roles with narrow entry paths, both sides reinforce a system that favors deepening focus while discouraging lateral movement.

Although the career funnel emerges as an adaptive structure, it is not the only equilibrium possible, nor is it necessarily an optimal one. In fact, it may carry long-term trade-offs across two critical dimensions for the individuals that follow it, namely in capability development and personal fulfillment. On one hand, highly specialized career trajectories risk locking individuals into early identity foreclosure, limiting their adaptive capacity when circumstances shift. Studies in career development (e.g., Ibarra, 2004) highlight how early commitment to a singular path can inhibit experimentation, skill diversity, and psychological mobility. Similarly, research on "career adaptability" (e.g. Savickas & Porfeli, 2012) underscores that the ability to pivot is a predictor of both performance and satisfaction in dynamic environments (Rudolph, Lavigne, Zacher 2017). Rigid career trajectories may also dampen personal fulfillment. Longitudinal research from the Harvard Study of Adult Development (e.g., Waldinger & Schulz, 2023) has shown that personal well-being correlates strongly with autonomy, relationships, and a sense of living on one's own terms, all factors that are rarely fully compatible with narrow occupational silos. The more specialized the career, the less room there may be to align work with evolving values, interests, or life priorities.

If the specialization trap can be suboptimal for the individual, might it also be suboptimal for the firm? While some research explores the link between cognitive diversity and innovation (Page, 2007), or between functional rotation and leadership development (Dragoni et al., 2009), surprisingly little has been written on the organization-level impacts of systems that over-index on deep expertise. This gap in the literature is telling, and suggests fertile ground for future inquiry into the "organizational cost of narrowness" which this paper now aims to address.

In this study, we investigate whether career breadth, defined as cross-functional, cross-sectoral, or otherwise domain-spanning professional experience, is associated with superior performance at the organizational level. Specifically, we focus on CEOs of publicly listed companies and examine the relationship between their career pathways and subsequent firm outcomes, as measured by long-term shareholder returns. Building on prior research in cognitive flexibility, learning transfer, and adaptive leadership, we ask whether broader career trajectories confer measurable advantages in navigating complex, volatile environments that most publicly listed companies represent. By operationalizing a novel "career range" metric and comparing high- and low-breadth profiles across industries, this study contributes empirical evidence to a growing conversation about whether the career funnel, while prevalent, may underperform more eclectic alternatives in contexts where adaptability is at a premium.

## 2 Literature review: the power of breadth

Whether it is athletic training or cognitive development, prior research has shown that diverse exposure and cross-contextual experience are often linked to superior performance outcomes. In fact, a growing body of evidence points to the advantages of breadth: exposure to a wider array of tasks, disciplines, or knowledge sets enhances adaptive capacity, creative problem solving, and performance transfer. These benefits emerge through mechanisms such as analogical reasoning, cognitive flexibility, and pattern recognition, particularly in dynamic environments that demand versatility. This section synthesizes these findings across physical, cognitive, and organizational research to build the case for breadth as a performance enhancer, motivating our inquiry into whether such effects might hold in the case of senior leadership and firm-level outcomes.

### 5.1 *Breadth and physical performance*

In sports science, the principle of varied training conditions has long been used to explain why athletes exposed to a broader range of stimuli often outperform those trained under narrowly consistent conditions. Chua et al. (2019) demonstrate that practice variability can lead to stronger retention and transfer of motor skills, with Caballero et al. (2024) refining the general results to accommodate for variability in the variability itself. Similarly, Seifert et al. (2013) argue that adaptability in elite sports emerges not from rote repetition but from dynamic interaction with unpredictable environments. These findings are consistent with the contextual interference theory (Battig, 1966; Schmidt & Bjork, 1992), which, posits that randomized practice schedules foster greater learning transfer than blocked or repetitive ones. Such performance improvements are not limited to elite athletes. Instead, Myer et al. (2011) found that youth athletes trained with a multilateral model that combines strength, agility, and coordination across a variety of sports developed superior neuromuscular control and reduced injury risk compared to early-specialized peers. These effects generalize to other domains where fine-tuned physical performance is critical. For example, in occupational training, Recent research (Elgendi et al., 2024) confirms that exposure to multiple varied simulation scenarios significantly improves first responders' adaptive capacity and clinical performance under stress, echoing earlier findings by Wilkerson et al. (2008) on the benefits of virtualized training for first responders. Together, these studies establish a strong precedent on how breadth in physical training improves performance.

### *2.2 Cognitive breadth and performance*

The cognitive sciences echo the findings above. Pattern recognition, a hallmark of expert judgment, has been shown to improve with exposure to varied and contrasting stimuli. Studies on cognitive flexibility (e.g. Spiro et al., 1990) argue that individuals trained with ill-structured, varied problems outperform those with narrowly defined tasks when confronted with novel situations. Wang et al. (2024) further confirm this across cultures, showing that high cognitive flexibility predicts academic performance even in tightly structured education systems, with the cognitive flexibility of the surrounding context playing an intermediating role. Interdisciplinary exposure has further been linked to long-term academic productivity, for example, Heiberger et al. (2021) found that even in fields with high returns to specialization, thematic consistency has diminishing returns while Liu et al (2025) find inter-disciplinary having a positive effect on funding performance. Extending this line of evidence, Prinzing and Vazquez (2025) demonstrate that students majoring in philosophy outperform peers from all other disciplines on standardized assessments of verbal and logical reasoning. These results suggest that the epistemic structure of philosophical training that emphasizes abstraction and open-ended inquiry, cultivates broadly transferable cognitive skills. The findings underscore the same insight of how cognitive performance seems to improve when individuals encounter diverse, cross-cutting material that challenges them reframe problems and synthesize information from multiple perspectives.

Lateral thinking refers to the capacity to approach problems from new angles by linking disparate concepts or reframing assumptions. Coined and popularized by Edward de Bono (1990), the term describes a mode of problem solving that departs from linear, stepwise reasoning in favor of generative, often counterintuitive idea formation. Unlike vertical thinking, which refines existing paths of logic, lateral thinking involves making novel connections that bypass conventional constraints, an ability strongly associated with innovation and creativity (De Bono, 1990). This approach aligns with Mednick's (1962) associative theory of creativity, which posits that individuals who can traverse "remote associative hierarchies" are more likely to produce original insights. Subsequent research has affirmed that breadth of experience plays a foundational role in enabling such cognitive jumps. For example, Duman et al. (2024) demonstrate that openness to inquiry, the disposition closely linked to intellectual curiosity and exploratory breadth, correlates with lateral thinking ability among higher education students. Complementary findings further reinforce the multidimensional value of lateral thinking. Related studies have connected different conceptualizations of lateral thinking to higher intelligence and academic achievement (Kumari & Aggarwal, 2012), effectiveness in technological education (Waks, 1997), improved speaking competence (Omran & Krebet, 2022), social innovation outcomes (Aydemir, 2021), empathetic tendencies (Arslan et al., 2022), and enhanced problem-solving skills (Yazgan, 2021) among other features. Together, these results suggest that fostering lateral thinking through diverse inputs, open inquiry, and exposure to varied knowledge domains, serves as a powerful mechanism by which cognitive and organizational performance can be elevated in certain contexts. Research also points to "question literacy" as a performance driver in leadership and knowledge work. The work of Loewenstein (1994) and Gillies et al. (2012) suggest that curiosity, a closely related concept, is both a learnable skill and correlated with higher cognitive performance.

Finally, Raviv, Lupyan, and Green (2022) synthesize decades of cognitive science to argue that variability in input is a central driver of learning generalization. Their review demonstrates that while exposure to more varied stimuli may initially slow learning, it consistently leads to deeper, more robust, and transferable understanding across contexts. This principle of validated across domains such as language and perception show that generalization at the level of inputs improves outputs, even if the path to mastery is less direct.

### *2.3. Synthesis: learning by patterns and contrasts*

One of the most enduring insights in the psychology of learning is that expertise is rooted in the recognition of meaningful patterns. Seminal work by Chase and Simon (1973) on chess expertise demonstrated that masters detect structured configurations or "chunks" of information that novices cannot. This perceptual advantage is not unique to chess; it extends across fields such as radiology (Kundel & Nodine, 1975), aviation (Endsley, 1995), and even business forecasting (Makridakis, 2009), where seasoned practitioners rely on deeply internalized schemas formed through repeated and varied experience. Pattern recognition, it seems, is not simply a function of repetition, but of structured diversity in inputs and what differentiates adaptive expertise from rote routine is the richness of contrasting exemplars through which concepts are learned and flexibly reapplied (e.g. Hatano & Inagaki, 1986; Chi et al., 1981).

The extant research suggests that breadth strengthens this process through two distinct mechanisms which we can synthesize as follows. First, by expanding the number and diversity of exposures, breadth increases the cognitive "inventory" available for constructing complex mental models, including the patterns to be matched. This includes examples closely aligned with the previously identified pattern, but also analogous and adjacent ones that add texture and depth to core schemas. We term this function *infill synthesis* as the process by which seemingly peripheral inputs enrich understanding within a central domain. Gentner and Markman (1997) highlight how analogical comparison fosters schema abstraction, allowing learners to extract common relational structures from superficially dissimilar cases. For example, Gick and Holyoak (1980) demonstrated that analogical reasoning improved markedly when individuals studied multiple structurally related problems rather than a single example, extended by Salas et al. (2012)

into the systematization of training and development in organizations. Infill synthesis is, at its core, a pattern-deepening function which provides variation that reinforces learning by introducing relevant permutations and contextual nuances.

Second, breadth facilitates sharper categorical distinctions by exposing learners to contrasting domains. What we refer to as the *silhouette effect*, analogous to visual edge detection in neuroscience, helps individuals define a category not just by its internal properties but by its contrast with adjacent categories. Murphy (2002) notes that categorization is inherently contrastive, suggesting that concepts gain coherence when learners understand where they end as much as where they begin. For example, novice clinicians taught to differentiate between closely related diagnoses (e.g., viral vs. bacterial infections) perform better than those taught via isolated exemplars (Norman et al., 2007). In organizational cognition, Kaplan and Tripsas (2008) show that executives who have worked across multiple strategic logics are more adept at reframing problems and resisting path-dependent reasoning. The silhouette mechanism enhances boundary recognition and minimizes the risk of misapplying heuristics formed in one context to another.

Together, *infill synthesis* and *silhouette contrast* offer a theoretical framework for understanding how breadth fuels adaptive expertise. The former broadens and deepens schema networks by layering analogical and adjacent variation, while the latter sharpens categorical resolution by delineating domain boundaries through contrast. These dual mechanisms support better pattern recognition and richer analogical reasoning when individuals face unfamiliar or ill-structured problems as well as in familiar contexts. Taken together, this body of research converges on a key insight of how variability and breadth function not as distractions from expertise, but as catalysts for deeper learning and broader application. They build the scaffolding necessary for individuals to flexibly recombine prior knowledge, adapt to novelty, and transcend the local optima, whether through lateral thinking, epistemological trespass or pure curiosity. Whether through the additive process of infill synthesis or the contrastive clarity of the silhouette effect, breadth appears to systematically strengthen the building blocks of expertise.

These findings offer a plausible bridge between individual learning and broader questions of leadership effectiveness. If variability and cross-domain exposure enhance individual cognition, can the same be said for professional breadth at the executive level? Might leaders who traverse multiple functional domains or sectors bring advantages in pattern discernment, strategic agility, or problem framing? These are the questions we turn to next. The following section introduces our empirical inquiry into whether such breadth correlates with measurable performance gains at the organizational level.

/// **Figure 2: Infill and silhouette effect** ///

*2.4 Breadth at the organizational level: teams and leaders*

The advantages of breadth explored in individual learning also raise a broader organizational question: might such variation similarly benefit firms? If varied experiences sharpen judgment, deepen cognitive flexibility, and support adaptive performance at the individual level, it is reasonable to hypothesize that these effects could scale up. Before addressing leadership directly, it is useful to revisit the nature of the firm itself.

Classic theories of the firm, most notably Coase's (1937) theory of transaction costs, conceptualize organizations as structured aggregations of individual agents brought together to reduce the inefficiencies of market transactions. From this foundational view, firms are inherently collections of individuals whose collective decisions and competencies shape outcomes. It follows that the composition, interactions, and capabilities of these individuals should meaningfully affect firm performance.

A growing body of literature supports this view. Research on team dynamics, for example, has shown that cognitive and functional diversity across organizational units affects decision quality, innovation rates, and adaptability (Page, 2007; van Knippenberg et al., 2004). These effect are expected to be particularly pronounced in knowledge-intensive environments where problem-solving benefits from exposure to contrasting perspectives (Horwitz & Horwitz, 2007; Joshi & Roh, 2009). Diversity is not only valuable at the team level, but may also hold importance at the level of individual leaders. In recent years, scholars have begun to explore how managerial experiences shape organizational outcomes. Maddux and Galinsky (2009), for instance, demonstrated that individuals with international experience exhibit higher integrative complexity and creativity. Their findings suggest that varied personal and professional experiences may enrich the decision-making capabilities of leaders, particularly in contexts requiring creative problem-solving and cross-cultural awareness. In organizational research, Dragoni et al. (2009) found that managers with cross-functional experience developed more sophisticated strategic thinking, and were more likely to attain upper-level executive roles. Their study underscores the idea that exposure to varied work domains fosters broader mental models and a more integrated understanding of the enterprise.

At the firm level, Custódio, Ferreira, and Matos (2013) analyzed U.S. CEOs and found that those whose work experience spanned multiple roles, industries, or geographies, received significantly higher compensation than their more specialized peers which in turn is typically tied to firm-level performance. These findings imply that the accumulated heterogeneity of a leader's experience can manifest in superior strategic outcomes. Despite the body of work in the area, an important gap remains. While numerous studies address the benefits of team-level diversity or the developmental advantages of career variety, relatively little research has explicitly examined the link between a leader's breadth of experience and the performance of the firm they lead. Put differently, given that we know that diverse teams work better, and that varied experience improves individual cognition, is there reason to believe that a career path marked by cross-domain exposure translates into better outcomes at the firm level?

This question lies at the heart of our present investigation. By focusing on the relationship between leader-level career breadth and firm performance, we aim to evaluate whether the benefits of breadth scale beyond individual cognitive gains and into the measurable performance dynamics of organizations.

## 3 Method, hypothesis and operationalization

### 5.1 *Defining and coding breadth of expertise of CEOs*

Building on the literature reviewed above, we hypothesize that the breadth of a leader's prior experience meaningfully predicts firm performance. This hypothesis rests on two complementary foundations. First, there is substantial evidence suggesting that breadth, defined as exposure to multiple domains, disciplines, or functional roles, improves individual-level cognitive performance, adaptability, and problem-solving capacity. Second, it is evident that the CEO of a firm is a central actor whose strategic decisions significantly shape organizational outcomes, and numerous studies underscore this link between leadership and performance. Bertrand and Schoar (2003), for example, show that CEO fixed effects explain a sizable portion of firm-level variation in profitability, investment, and strategy, even after controlling for industry and firm characteristics. In other words, who leads a firm matters in terms of bottom-line results. If leadership matters, and breadth enhances leadership cognition, then it follows that CEOs with broader experience sets may position their firms for superior performance, particularly in complex, volatile environments.

To test this hypothesis, we conceptualize breadth in this context along two dimensions: educational exposure and professional experience. For each CEO, we build a composite index using information drawn from public resume data (e.g., annual reports, investor disclosures, press releases, and public databases).

Educational breadth is assessed by analyzing the range and heterogeneity of the CEO's formal academic qualifications. This includes the number of degrees attained, the disciplinary spread of those degrees, and the diversity of institutions attended. Degrees are classified into broad disciplinary clusters such as humanities, social sciences, STEM, business and finance, and legal-regulatory fields based on the Digital Commons Disciplines Taxonomy. For example, a CEO holding degrees in both engineering and philosophy, would be considered more educationally broad than one with multiple degrees in economics alone. While the raw count of degrees provides a baseline, the key metric is the number of distinct knowledge domains represented, reflecting the variety of epistemic exposure.

**Table 1. Categorization of educational disciplines**

| Field Cluster | Sample Disciplines |
|---|---|
| Humanities | History, Philosophy, Literature |
| Social Sciences | Economics, Political Science, Sociology |
| STEM | Engineering, Physics, Computer Science |
| Business and Finance | Accounting, Finance, Management |
| Legal and Regulatory | Law, Public Policy |

Professional breadth is operationalized by examining the executive's career history across industries and functional domains. Each distinct role in a CEO's career is treated as a discrete episode of learning and domain engagement. The more sectors and functions a leader has traversed, the greater the assumed breadth. Functional domains are coded into categories such as finance, operations, marketing, technology, and corporate strategy, allowing for a structured comparison of role diversity following NAICS (Table 2).

**Table 2. Domains and Industry sectors**

**Functional Domains (Role Types)**

| Code | Functional Domain | Description |
|---|---|---|
| FIN | Finance | Corporate finance, treasury, accounting, FP&A, investment roles |
| OPS | Operations | Supply chain, manufacturing, logistics, process optimization |
| MKT | Marketing & Sales | Brand, product marketing, customer insights, sales strategy |
| TEC | Technology & Engineering | IT, software development, systems engineering, R&D |
| STR | Strategy & Corporate Dev | M&A, business development, strategic planning |
| HRM | People & Organization | HR, org design, talent management, DEI initiatives |
| LEG | Legal, Risk & Compliance | Legal affairs, corporate governance, regulatory strategy |
| GEN | General Management | Business unit leadership, P&L roles, COO-type roles |

**Industry Sectors (Domain Context)**

| Code | Industry Sector | Description |
|---|---|---|
| CON | Consumer Products & Retail | FMCG, e-commerce, apparel, food & beverage |
| FIN | Financial Services | Banking, insurance, asset gmt., fintech |
| HLT | Healthcare & Life Sciences | Pharma, medtech, hospitals, biotech |
| TEC | Technology & Software | SaaS, hardware, IT services, platforms |
| IND | Industrial & Manufacturing | Heavy industry, capital goods, automotive |
| ENE | Energy & Utilities | Oil & gas, renewables, grid infrastructure |
| COM | Communications & Media | Telecom, digital media, publishing |
| GOV | Public Sector & Education | Government, NGOs, education, multilateral organizations |
| LOG | Transport & Logistics | Airlines, shipping, supply chain, distribution |
| PRO | Professional Services | Consulting, law, accounting, design, advisory |

We compute a composite score for each CEO's educational and professional breadth based on this schema (Table 3). While the primary focus in this initial model is on the count and type of transitions, we note a limitation of this approach: it does not yet incorporate depth of experience within a given role or field. Consequently, a short tenure in a novel function is treated equivalently to a long-term role, which may overstate effective exposure. Future refinements may address this limitation by weighting experiences by duration or incorporating performance evaluations within each role where data is available.

This operational framework enables us to quantify breadth in a structured and replicable way. By translating qualitative career histories into standardized metrics, we create a dataset amenable to statistical analysis. The next section describes the data sources, collection protocols, and sample construction used to test the hypothesis. We also outline the performance metrics employed to assess firm outcomes, including return on assets, revenue growth, and innovation indicators, in relation to the breadth indices developed here.

**Table 3. Weighted schema**

| **Dimension** | **Metric** | **Description** | **Scoring Method** | **Limitations** |
|---|---|---|---|---|
| Educational breadth | Number of degrees | Count of distinct academic degrees attained (e.g., B.A., M.Sc., Ph.D.) | Raw count | Does not account for degree length or quality |
| Educational breadth | Field diversity score | Count of distinct academic disciplines (e.g., History, Finance, Engineering) | Raw count | Treats closely related disciplines as distinct (e.g., Econ vs. Finance) |
| Professional breadth | Number of unique roles | Total number of distinct job titles or leadership positions held | Raw count | Short and long tenures treated equally |
| Professional breadth | Functional domain diversity | Count of unique domains (e.g., Finance, Operations, Marketing, Strategy, etc.) | Raw count | Overlooks nuance between sub-functions (e.g., M&A vs. broader strategy roles) |
| Professional breadth | Industry sector diversity | Count of distinct industries worked in (e.g., Tech, Finance, Consumer Goods) | Raw count | Does not weight by sector size or complexity |
| Professional breadth | Cross-sector transitions | Count of clear transitions between industries or major functions | Raw count of transitions | Ignores depth of adaptation or learning within the transition |

*3.2 Modeling CEO breadth and firm performance*

To empirically assess the relationship between CEO breadth and firm-level performance, we constructed a dataset consisting of the 650 largest publicly traded companies by market capitalization as of 2024. This sample offers both scale and sectoral diversity, enabling us to draw statistically robust inferences while accounting for industry-specific dynamics and redundancies to ensure our study covers the 500 largest publicly traded companies. The central question guiding this analysis is whether the professional and educational breadth of a firm's chief executive correlates with superior stock performance following their appointment.

We define a CEO's "event window" as the thirty-six months following the quarter of their formal assumption of office. This window balances the need for sufficient post-appointment performance data with the practical constraints of executive turnover and market volatility. Our primary outcome variable is firm stock return, which we measure in terms of abnormal returns that deviate from expected performance benchmarks. To provide a rigorous baseline, we compute excess returns against three distinct comparators. First, we calculate market-adjusted returns by subtracting the performance of a broad-based index (the S&P 500) from the firm's own return over the same period. Second, we compute peer-adjusted returns by benchmarking each firm against the median return of its sector, defined by two-digit GICS industry classifications. This accounts for industry-specific shocks or booms. Third, we consider a firm's own pre-appointment performance trajectory by measuring deviations from its average returns over the thirty-six months preceding the CEO's appointment. This "momentum-adjusted" metric allows us to control for firm-specific trends and mean reversion effects.

The core explanatory variable is the CEO's breadth index, a composite measure described in the prior section, which aggregates educational and professional transitions across domains. CEOs are sorted into quintiles based on their breadth scores, ranging from the most specialized to the most broadly experienced. We then evaluate whether firms led by CEOs in the higher breadth quintiles systematically outperform those led by more specialized leaders, using the three return metrics described above. To isolate the effect of CEO breadth from other firm and leader characteristics, we introduce a set of control variables. These include firm size, leverage, R&D intensity, board composition, and CEO-specific variables such as age, tenure, and compensation structure. We employ an OLS regression framework with industry-clustered robust standard errors and include fixed effects for year and sector to further mitigate confounding influences. In addition, we conduct robustness checks using propensity score matching and lagged dependent variable models to validate the stability of our findings.

Through this modeling strategy, we aim to evaluate whether career breadth functions as a predictive variable for firm performance, under real-world capital market conditions. In the next section, we describe the data sources, protocols for resume coding and breadth index construction, and the detailed steps taken to build and validate our dataset. This groundwork sets the stage for empirical testing of the hypothesis that broader career paths in executive leadership yield measurable value creation for shareholders.

**Table 4. Firm analysis**

| **Component** | **Specification** |
| --- | --- |
| **Sample** | 650 largest public firms by market cap (2023) |
| **Event window** | 36 months following CEO appointment |
| **Breadth variable** | Composite index of educational + professional breadth (quintiles) |
| **Dependent variables** | Abnormal stock returns (peer-adjusted, market-adjusted, momentum-adjusted) |
| **Benchmarks** | Firm’s historical returns, industry median, and S&P 500 index |
| **Control variables** | Firm size, industry fixed effects, R&D intensity, leverage, CEO age, tenure, pay |
| **Statistical method** | OLS regression with robust standard errors; industry-clustered fixed effects |
| **Robustness checks** | Propensity score matching; lagged dependent variable models |

We detail data sources, variable construction and analytical procedures in Technical Appendix 1.

## 4 Results

// RESULTS UNDER REVISION //

## 5. Discussion

// RESULTS UNDER REVISION //